\documentclass[aps,prl,twocolumn,superscriptaddress,citeautoscript,amssymb,reprint,showpacs,floatfix]{revtex4-1}

\usepackage[ansinew]{inputenc}
\usepackage[T1]{fontenc}
\usepackage{amsmath}
\usepackage[normalem]{ulem}
\usepackage[english]{babel}
\usepackage{graphicx}
\usepackage{color}
\usepackage{natbib}
\usepackage{tabularx}

\begin{document}
\author{J. F. Landaeta}
\email[]{Javier.Landaeta@cpfs.mpg.de}
\affiliation{Max Planck Institute for Chemical Physics of Solids, 01187 Dresden, Germany}

\author{A. M. Leon}
\affiliation{Max Planck Institute for Chemical Physics of Solids, 01187 Dresden, Germany}

\author{S. Zwickel}
\affiliation{Technical University Munich, Physics department, 85748 Garching, Germany}

\author{T. L\"uhmann}
\affiliation{Max Planck Institute for Chemical Physics of Solids, 01187 Dresden, Germany}

\author{M. Brando}
\affiliation{Max Planck Institute for Chemical Physics of Solids, 01187 Dresden, Germany}

\author{C. Geibel}
\affiliation{Max Planck Institute for Chemical Physics of Solids, 01187 Dresden, Germany}	

\author{E.-O. Eljaouhari}
\affiliation{Technische Universit\"{a}t Braunschweig, Institut f\"{u}r Mathematische Physik, Mendelssohnstra{\ss}e 3, 38106 Braunschweig, Germany}
\affiliation{Universit\'{e} de Bordeaux, CNRS, LOMA, UMR 5798, 33400 Talence, France}

\author{H. Rosner}
\affiliation{Max Planck Institute for Chemical Physics of Solids, 01187 Dresden, Germany}

\author{G. Zwicknagl}
\affiliation{Max Planck Institute for Chemical Physics of Solids, 01187 Dresden, Germany}
\affiliation{Technische Universit\"{a}t Braunschweig, Institut f\"{u}r Mathematische Physik, Mendelssohnstra{\ss}e 3, 38106 Braunschweig, Germany}

\author{E. Hassinger}
\email[]{Elena.Hassinger@cpfs.mpg.de}
\affiliation{Max Planck Institute for Chemical Physics of Solids, 01187 Dresden, Germany}
\affiliation{Technical University Munich, Physics department, 85748 Garching, Germany}

\author{S. Khim}
\email[]{Seunghyun.Khim@cpfs.mpg.de}
\affiliation{Max Planck Institute for Chemical Physics of Solids, 01187 Dresden, Germany}

\title{Conventional type-II superconductivity in locally non-centrosymmetric LaRh$_2$As$_2$ single crystals}
\date{\today}

\begin{abstract}
We report on the observation of superconductivity in LaRh$_2$As$_2$, which is the analogue without $f$-electrons of the heavy-fermion system with two superconducting phases CeRh$_2$As$_2$\cite{khim2021}. A zero-resistivity transition, a specific-heat jump and a drop in magnetic ac susceptibility consistently point to a superconducting transition at a transition temperature of $T_c = 0.28$\,K. The magnetic field-temperature superconducting phase diagrams determined from field-dependent ac-susceptibility measurements reveal small upper critical fields $\mu_{\mathrm{0}}H_{c2} \approx 12$\,mT for $H\parallel ab$ and $\mu_{\mathrm{0}}H_{c2} \approx 9$\,mT for $H\parallel c$. The observed $H_{c2}$ is larger than the estimated thermodynamic critical field $H_c$ derived from the heat-capacity data, suggesting that LaRh$_2$A$s_2$ is a type-II superconductor with Ginzburg-Landau parameters $\kappa^{ab}_{GL} \approx 1.9$ and  $\kappa^{c}_{GL}\approx 2.7$.
The microscopic Eliashberg theory indicates superconductivity to be in the weak-coupling regime with an electron-phonon coupling constant $\lambda_{e-ph} \approx 0.4$. Despite a similar $T_c$ and the same crystal structure as the Ce compound, LaRh$_2$As$_2$ displays conventional superconductivity, corroborating the substantial role of the 4$f$ electrons for the extraordinary superconducting state in CeRh$_2$As$_2$.
\end{abstract}

\maketitle
\section{Introduction}
\begin{figure}[t]
	\centering
	\includegraphics[width=0.9\linewidth]{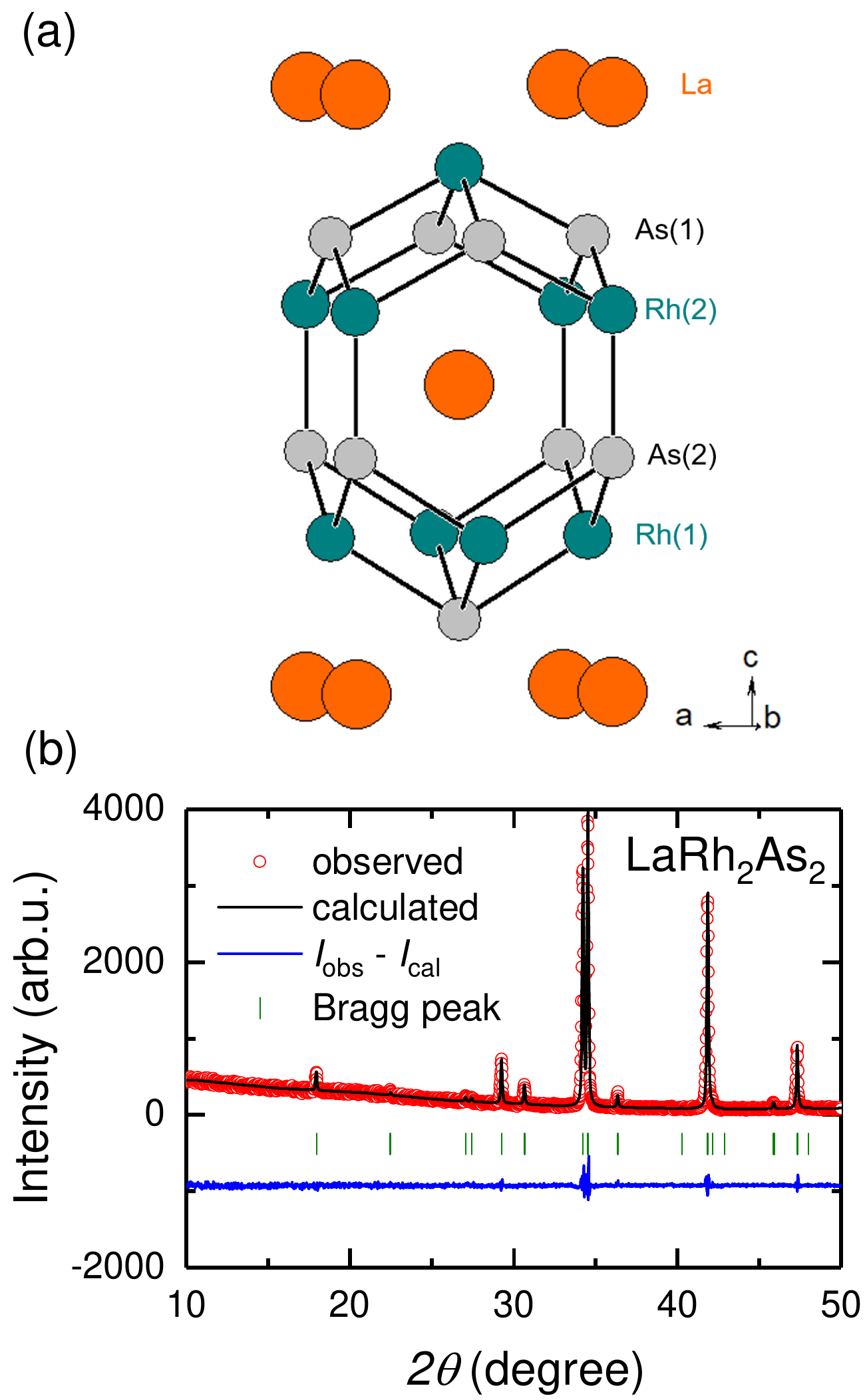}
	\caption{(a) Crystal structure of LaRh$_2$As$_2$ (b) Powder x-ray diffraction pattern and Rietveld refinement results
	}
\label{fig:xrd_strcuture}
\end{figure}

Recently, two-phase superconductivity has been reported for the heavy-fermion compound CeRh$_2$As$_2$ \cite{khim2021} with highly anisotropic critical fields $H_{c2}$ and one of the highest $H_{c2}/T_c$ values. The unique phase diagram and the large critical field value seem to originate in the crystal structure. Unlike many unconventional '122'-superconductors which crystallize in the ThCr$_2$Si$_2$-type structure, such as the celebrated superconductors CeCu$_2$Si$_2$ \cite{Steglich1979}, URu$_2$Si$_2$ \cite{Palstra1985}, and doped BaFe$_2$As$_2$ \cite{Rotter2008}, CeRh$_2$As$_2$ forms the CaBe$_2$Ge$_2$-type structure. In this structure, due to an exchange of Rh and As positions in half of the unit cell, inversion symmetry is broken locally at the Ce site (shown in Fig. \ref{fig:xrd_strcuture}(a)), while keeping an overall inversion symmetry.

As widely discussed in non-centrosymmetric superconductors, the (local or global) inversion symmetry breaking gives rise to asymmetric spin-orbit coupling (ASOC) \cite{Bauer2012,Smidman2017}. This induces an enhancement of $H_{c2}$ for specific field orientations for which the magnetic field is much less effective in polarizing the spins of the Cooper pairs. As a result, the Pauli limiting becomes ineffective, allowing for higher $H_{c2}$. However, due to the global inversion symmetry in CeRh$_2$As$_2$, the sign of ASOC is opposite for each Ce sublayer in the unit cell and the superconducting gap can either follow the sign change of the ASOC or not, leading to odd- and even parity superconducting states \cite{Yoshida2012,Maruyama2012,Sigrist2014,Zhang2014}. This additional parity degrees of freedom lead to the field-induced phase transition within the superconducting state and the enhanced $H_{c2}$ in the high-field phase \cite{khim2021}. This suggests that the lack of local inversion symmetry and ASOC are key ingredients towards the two-phase superconductivity observed in CeRh$_2$As$_2$\cite{khim2021}. In addition, the inherent nonsymmorphic symmetry might lead to a topological crystalline superconducting state \cite{Wang2017,Nogaki2021}.

The aim of this study is to characterize the La analogue, LaRh$_2$As$_2$ without 4$f$ electrons, reported to crystallize in the same structure, but which has not been studied down to low temperatures \cite{Madar1987}. 
In doing so, we are able to investigate the influence of the crystal structure itself and the role of the 4$f$ electrons that are absent in the La compound. Indeed, in non-centrosymmetric heavy-fermion superconductors it was found that broken inversion symmetry is not the only requirement for unconventional superconductivity \cite{Ribeiro2009}. Rather the 4$f$ electrons must actually play a crucial role since all unconventional superconducting properties have not been observed in their La analogues \cite{Settai2008,palazzese2018strong}.

We report on the single crystal growth of LaRh$_2$As$_2$ and resistivity, specific heat and magnetic ac-susceptibility properties down to low temperatures. We observe a superconducting transition at $T_c$ = 0.28\,K and determine the superconducting magnetic field $H$-temperature $T$ phase diagram of $H_{c2}$. Estimated superconducting parameters point to type-II superconductivity. An estimation of the electron-phonon coupling based on the Eliashberg theory points to a weak-coupling superconductor. Comparing the Sommerfeld coefficient deduced from a first principle band structure calculation and the experimental value confirms the weak electron-phonon coupling. We find that - despite a similar $T_c$ and the same crystal structure - LaRh$_2$As$_2$ displays conventional superconductivity, corroborating the substantial role of the Ce-4$f$ electrons for unconventional superconductivity in CeRh$_2$As$_2$.

\section{Methods}
\subsection{Experimental Methods}
Single crystals of LaRh$_2$As$_2$ were grown by the Bi-flux method.
Pure elements in the ratio La:Rh:As:Bi = 1:2:2:30 were placed in an alumina crucible and subsequently sealed in a quartz tube filled with 300\,mbar argon. The crucible was heated to 1150 $^{\circ}$C for 4 days and slowly cooled down to 700\,$^{\circ}$C for a week by slowly lowering the crucible in a vertical furnace. Grown single crystals were extracted by etching the Bi flux in diluted nitric acid solution. The crystal structure was analyzed by powder x-ray diffraction measurements.

The heat-capacity measurements were carried out down to 0.5\,K using the relaxation-time method in a Quantum Design Physical Property Measurements System (PPMS). A customized compensated heat-pulse calorimeter was used between 0.04 and 4\,K in a dilution refrigerator \cite{Wilhelm2004}. For the low temperature measurements, the nuclear contribution was removed with the same procedure as described in \cite{khim2021}. At low temperatures, the contributions to the heat-capacity are given by $C=C_{n}+C_{el}+C_{ph}= \frac{\alpha}{T^2} +\gamma_0 T + \beta T^3$. Where, $C_{el}$ + $C_{ph}$ are the electronic and phonon contributions and $\alpha$ is the proportionality factor in the nuclear Schottky specific heat $C_n$. In the case of LaRh$_2$As$_2$ as well CeRh$_2$As$_2$, the nuclear contribution only comes from As atoms at the two crystallographic sites. From NQR measurements \cite{Kibune2022}, we know that the quadrupolar splittings $\Delta/h$ for As(1) and As(2) are 31.1 MHz and 10.75 MHz respectively, where the nuclear spin of the As atom is $I=3/2$ with 100 \% abundance. With the expression of the high-temperature term of the Schottky molar heat capacity at zero magnetic field $\alpha=R/4(\Delta/k_B)^2$\cite{Hagino1994}, where $R$ is the gas constant, we obtain the values of the $\alpha$ parameter for each As site, where the total contribution is $\alpha=5.196\times10^{-6}$ JK/mol. For temperature below 1K, the phonon contribution is very small and can be ignored from the contributions to the specific heat. Finally, we can use $C/T=\alpha/T^3+\gamma_0$ to remove the nuclear contributions, where $\gamma_0=10.5$ mJ/molK$^2$.

The magnetic ac-susceptibility was measured using a homemade set of compensated pick-up coils of 2\,mm length and 6000\,turns each. The inner and outer diameter was 1.8\,mm and 5\,mm, respectively. A superconducting modulation coil produced the excitation field of 40\,$\mu$T at 1127\,Hz. The output signal of the pick-up coils was amplified using a low-noise amplifier SR560 from Stanford Research Systems.

Our setup uses a National Instruments 24 bits PXIe-4463 signal generator and 24 bits PXIe-4492 oscilloscope as data acquisition system with digital lock-in amplification. The ac-susceptibility measurements were performed for an applied field up to 15\,T parallel and perpendicular to the tetragonal $c$ axis using a small single crystal of a volume of $\sim$ 750\,$\mu$m$^3$ down to 35 \,mK in a MX400 Oxford dilution refrigerator. For the magnetic field dependence of the ac-susceptibility measurements, the remnant field of the superconducting magnet was corrected. The remnant magnetic field was about $\pm$ 2 mT when sweeping the field between - 25 mT to 25 mT as done here. It is important to remark that we started the measurement with a demagnetized magnet.

For the resistivity measurements, a standard 4-point method was employed with current and voltage contacts along a line perpendicular to the $c$ axis using an excitation current of 1\,nA. The four contacts were made using 25\,$\mu$m diameter gold wires on a sample with silver paste (DuPont 4922N). The signal was amplified by a low-temperature transformer with a winding ratio of 1:100 and the output of the transformer was measured using a PXI lock-in setup at 113\,Hz.

\subsection{Computational Details}

Electronic structure calculations were performed applying density functional theory (DFT) \cite{Hohenberg1964,Kohn1965} as implemented in the Vienna \textit{ab initio} simulation package (VASP) \cite{Kresse1996} and the full-potential local-orbital code FPLO \cite{fplo}. Both the local density approximation (LDA; in the parametrization of Perdew and Wang \cite{PW92}) and the generalized gradient approximation (GGA; in the parametrization of Perdew-Burke-Ernzerhof \cite{perdew1996generalized}) were used. 

In the Vienna calculations a Monkhorst Pack grid of (26$\times$26$\times$16) was employed, the valence of the atoms are La: 5$s^{2}$5$p^{6}$5$d$6$s^{2}$, Rh: 5$s$4$d^{8}$, and As: 4$s^{2}$4$p^{3}$. The structural optimization is performed within a force convergence of at least 10$^{-3}$ eV/\AA\; for each atom, and a plane-wave energy cutoff of 500\,eV in order to relax the internal positions and the lattice parameters. Phonon spectra were obtained by using the PHONOPY \cite{pho} program with the interatomic force constants which are calculated by VASP using the linear response method based on density-functional perturbation theory. The calculations were performed using a supercell with 80 atoms (2$\times$2$\times$2 unit cell), for which the total energy is converged using a 8$\times$8$\times$6 $\textit{k}$-mesh and a 500\,eV plane wave cutoff.

In the FPLO calculations, the standard basis set and a $24\times24\times24$ $\textit{k}$-mesh were applied. The spin-orbit coupling (SOC) was taken into account non-perturbatively, solving the four component Kohn-Sham-Dirac equation.

The lattice parameters were taken from the experiment (see below).
Relaxation of the internal positions was performed minimizing the total energy applying GGA.The results were robust comparing both codes and both functionals employed. For a given structure, the respective electronic structures did not show significant differences.

\section{Results and discussion}
\subsection{Crystal structure}

Figure~\ref{fig:xrd_strcuture}(b) shows powder x-ray diffraction pattern and Rietveld refinement results with the tetragonal CaBe$_2$Ge$_2$-type structure (Fig.~\ref{fig:xrd_strcuture}(a)). We neither found any evidence for an additional phase, nor evidence for a ThCr$_2$Si$_2$-type structure. The accuracy of the spectrum limits the possible amount of interchange of As and Rh to a few percent. The determined lattice parameters $a$ = 4.313\,\AA\, and $c$ = 9.879\,\AA\,are comparable to the previous report \cite{Madar1987}. Internal atomic positions were obtained by density functional calculations (see subsection "Electronic structure calculations") and agree well with the closely related LaIr$_2$As$_2$ compound \cite{Pfannenschmidt2012}. Comparing with CeRh$_2$As$_2$ ($a$ = 4.280\,\AA \, and $c$ = 9.861\,\AA) $a$ and $c$ are 0.8 and 0.2\,\% larger, respectively, explained by the absence of the Ce-4$f$ electron. 

The structure has some similarity with the ThCr$_2$Si$_2$-type structure, since in half of the unit cell, the Rh(1) and As(2) in the CaBe$_2$Ge$_2$-type structure have the same layer order as the Rh and As in the centrosymmetric ThCr$_{2}$Si$_{2}$-type structure. However, the other Rh-As layer has a completely inverted arrangement so that the As(1) atom is now at the center of the layer. Remarkably, the in-plane bond distance As(1)-Rh(2) of 2.45 \AA~in the "inverted" layer is significantly shorter than the in-plane Rh(1)-As(2) distance of 2.54\,\AA~in the "normal" layer.
The inter-plane distance between the Rh(2) and As(2) atoms with 2.41\,\AA\ is the shortest Rh-As bond in the structure, small enough to form an additional strong bond. The bonding character between these Rh(2) and As(2) atoms is evidenced in the sister compound LaNi$_{2}$As$_{2}$, which crystallizes in both 122-polymorphs: the $c$-lattice parameter in the CaBe$_{2}$Ge$_{2}$-type structure is substantially shorter (9.47\,\AA) than that in the ThCr$_{2}$Si$_{2}$-type structure (9.93\,\AA) where this strong bond is absent \cite{Ghadraoui1988}.

\subsection{Normal state properties}

\begin{figure}[t]
	\centering
	\includegraphics[width=0.9\linewidth]{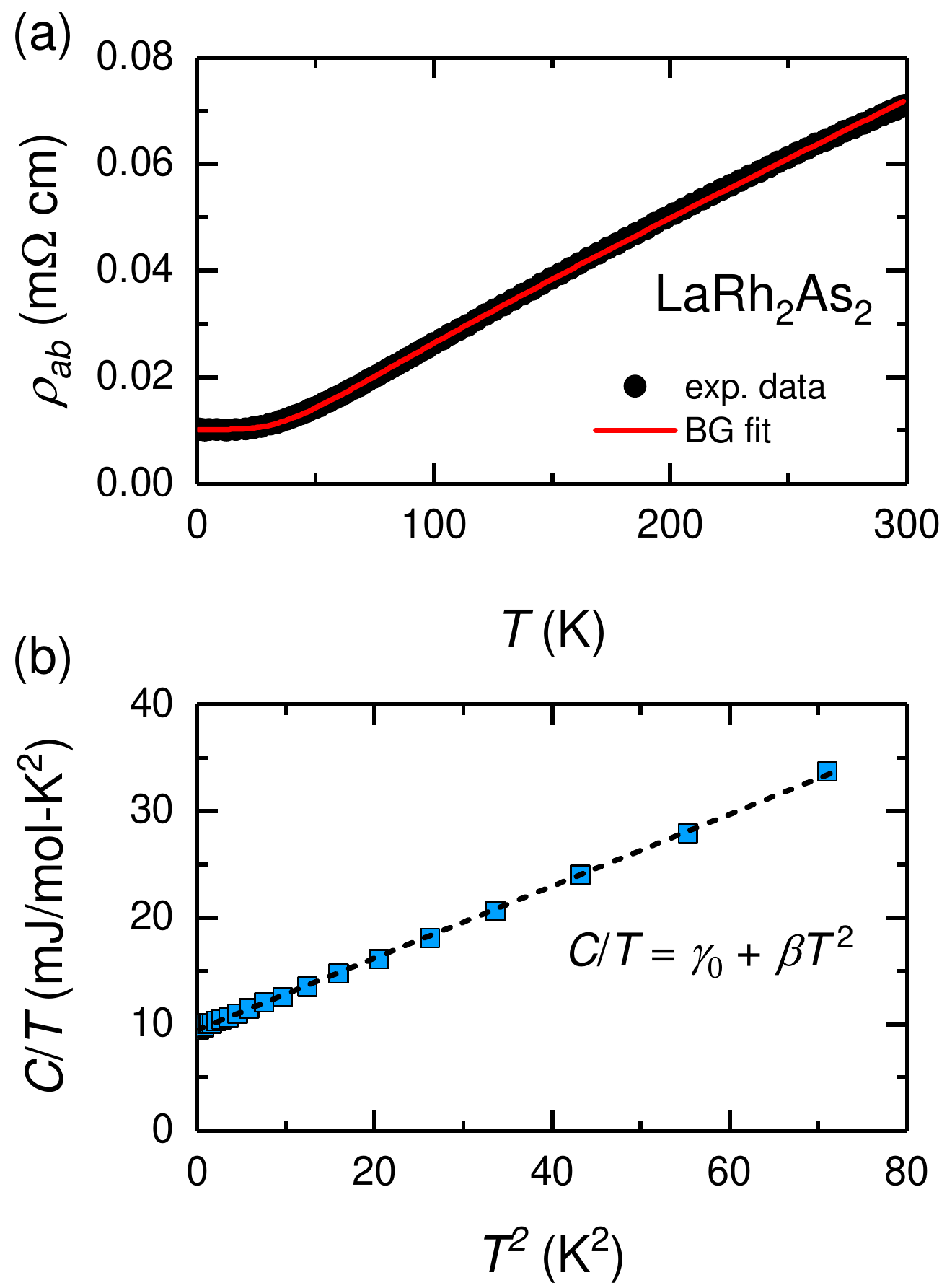}
	\caption{(a) Temperature-dependent resistivity (circles) and the Bloch-Gr\"{u}neisen fit (line). (b) Plot of $C$/$T$ as a function of $T^2$. The dotted line denotes the fit to the linear function $C/T$ = $\gamma_0$ + $\beta T^2$, where $\gamma_0$ is the Sommerfeld coefficient and the $\beta$ is the phonic specific-heat coefficient. 
	}
\label{fig:rho_T}
\end{figure}

The high-temperature resistivity $\rho(T)$ is shown in Fig.\,\ref{fig:rho_T}(a).
The room-temperature resistivity is $\rho_{300 K}$ = 0.079\,m$\Omega$cm and $\rho(T)$ monotonically decreases with lowering temperatures. Due to the absence of a magnetic element in LaRh$_2$As$_2$, the temperature dependence of $\rho(T)$ is likely attributed to electron-phonon scattering at high temperatures. Accordingly, the Bloch-Gr\"{u}neisen formula is applied to fit the $\rho(T)$ data, which is written as 

\begin{align}
    \rho (T) &= \rho_0 +
        \rho_{BG}(T) \nonumber \\
        &=\rho_0 + C_{BG} \left( \frac{T}{\Theta_R} \right)^{5} \int_{0}^{\Theta_R / T} \frac{x^5}{(e^x - 1)(1 - e^{-x})} dx,
\end{align}

where $\rho_0$ is the residual resistivity originating from imperfections of a crystalline lattice, and $\Theta_R$ is the characteristic temperature compatible with the Debye temperature.
The fitted curve reproduces the experimental data up to 300\,K (Fig. \ref{fig:rho_T}(a)) with $\Theta_R = 230$\,K, $C_{BG} = 0.019$\,m$\Omega$cm, and $\rho_0 = 0.010$\,m$\Omega$cm. The residual resistivity ratio (RRR = $\rho_{300 K}$ / $\rho_0$) is about 7.

The temperature dependence of the specific heat ($C$) is shown in Fig. \ref{fig:rho_T}(b), which follows the relation of $C/T = \gamma_0 + \beta T^2$ up to $\approx 9$\,K, where $\gamma_0$ is the Sommerfeld coefficient and $\beta$ is the phononic specific-heat coefficient. $\gamma_0$ is 9.4 mJ/mol-K$^2$ and $\beta$ is 0.34\,mJ/mol-K$^4$. The Debye temperature given by $\beta$ based on the relation $\theta_D = (12\pi^4  RN/5\beta)^{1/3}$ is 300\,K, where $R$ is the ideal gas constant and $N=5$ is the number of atoms in the primitive unit cell. This is larger than $\Theta_R$ determined from the Bloch-Gr\"{u}neisen fit of the resistivity data. This difference is not surprising: The low temperature specific heat probes the phonon density at very low energies, which is only determined by the velocity of the low-$Q$ acoustic modes, which scales with the inverse mass of all elements forming the compound. Instead the increment in $\rho(T)$ at higher temperature is sensitive not only to the acoustic modes, but to the optical modes, especially the lower ones whose energy scales with the inverse mass of the heaviest element.

\begin{figure}[t]
	\centering
	\includegraphics[width=0.9\linewidth]{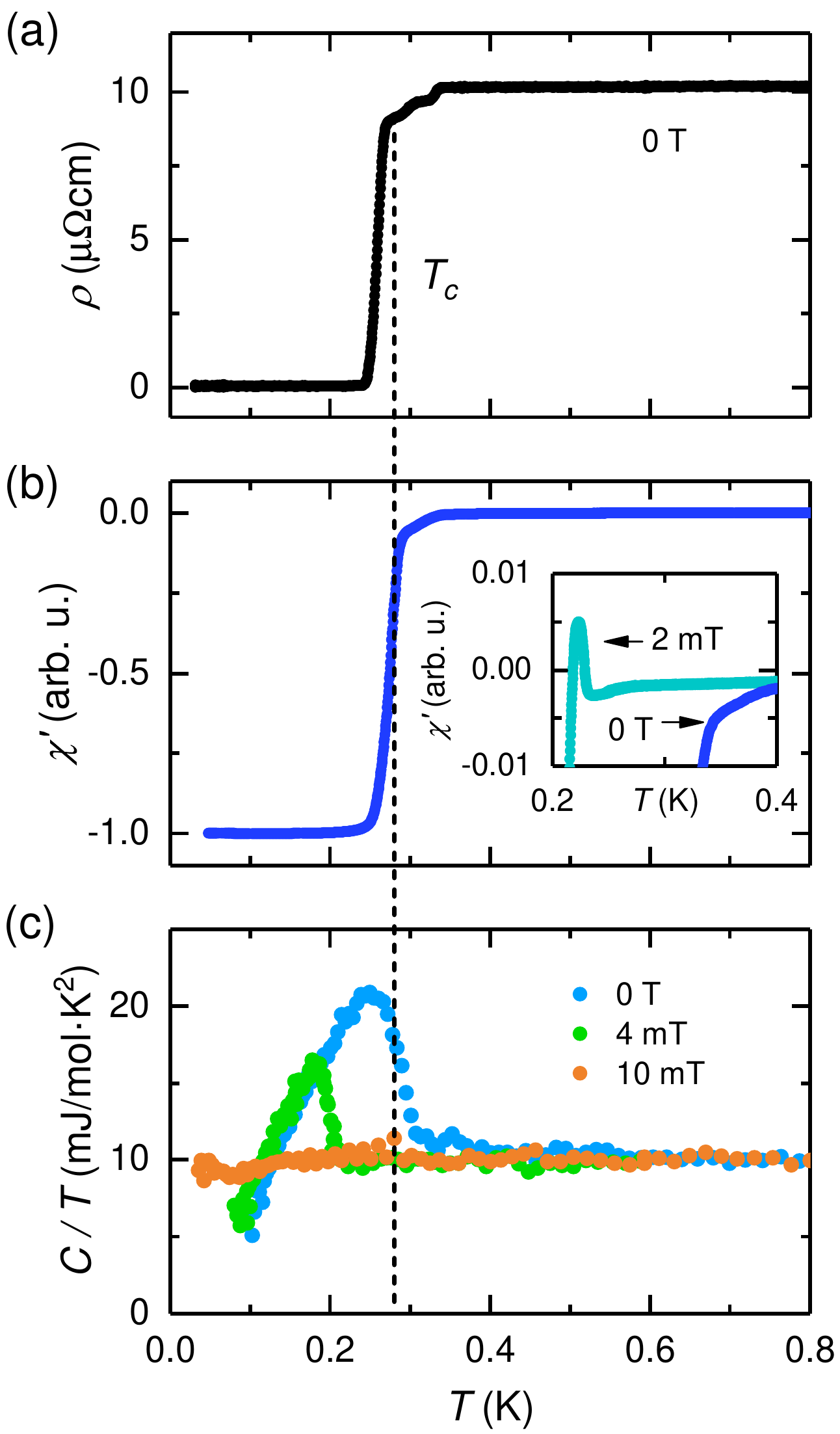}
	\caption{(a) Temperature dependence of resistivity $\rho(T)$ at zero field. (b) Temperature dependence of real-part of ac-susceptibility $\chi^{'}$($T$) at zero field. The inset  presents additional 2\,mT data obtained for the field perpendicular to the $c$ axis. (c) Temperature dependence of specific heat $C/T(T)$ measured at zero, 4, and 10\,mT along the $c$ axis. The vertical dotted line denotes the superconducting transition temperature determined using the equal entropy criterion, $T_c$ = 0.28\,K.}
\label{fig:Figure_3}
\end{figure}

\subsection{Superconducting phase transition}

Figure \ref{fig:Figure_3} highlights the experimental evidence for the superconducting phase transition.
A clear resistivity drop appears below $T_c$ = 0.28\,K (Fig. \ref{fig:Figure_3}(a)). At the same temperature, a drop in the real part of the ac-susceptibility $\chi'$ indicates the onset of diamagnetic shielding (Fig. \ref{fig:Figure_3}(b)).
Finally, the clear jump in $C/T$ (Fig. \ref{fig:Figure_3}(c)) evidences the bulk nature of the transition. An external field of 10\,mT along the $c$ axis completely suppresses the $C/T$ anomaly, indicating that $H_{c2}$ is of the order $\sim$\,10 mT. The ratio $\Delta C/\gamma_0T_c = 1.15$ is slightly smaller than the expected BCS value. 
This might be due to multiband superconductivity, or some fraction of the sample being non-superconducting, or to some additional contribution in the normal state specific heat. A small drop in $\rho(T)$ occurs well before the bulk $T_c$, likely due to spurious superconductivity caused by the presence of crystal imperfections.

\begin{figure}[t]
	\centering
	\includegraphics[width=0.9\linewidth]{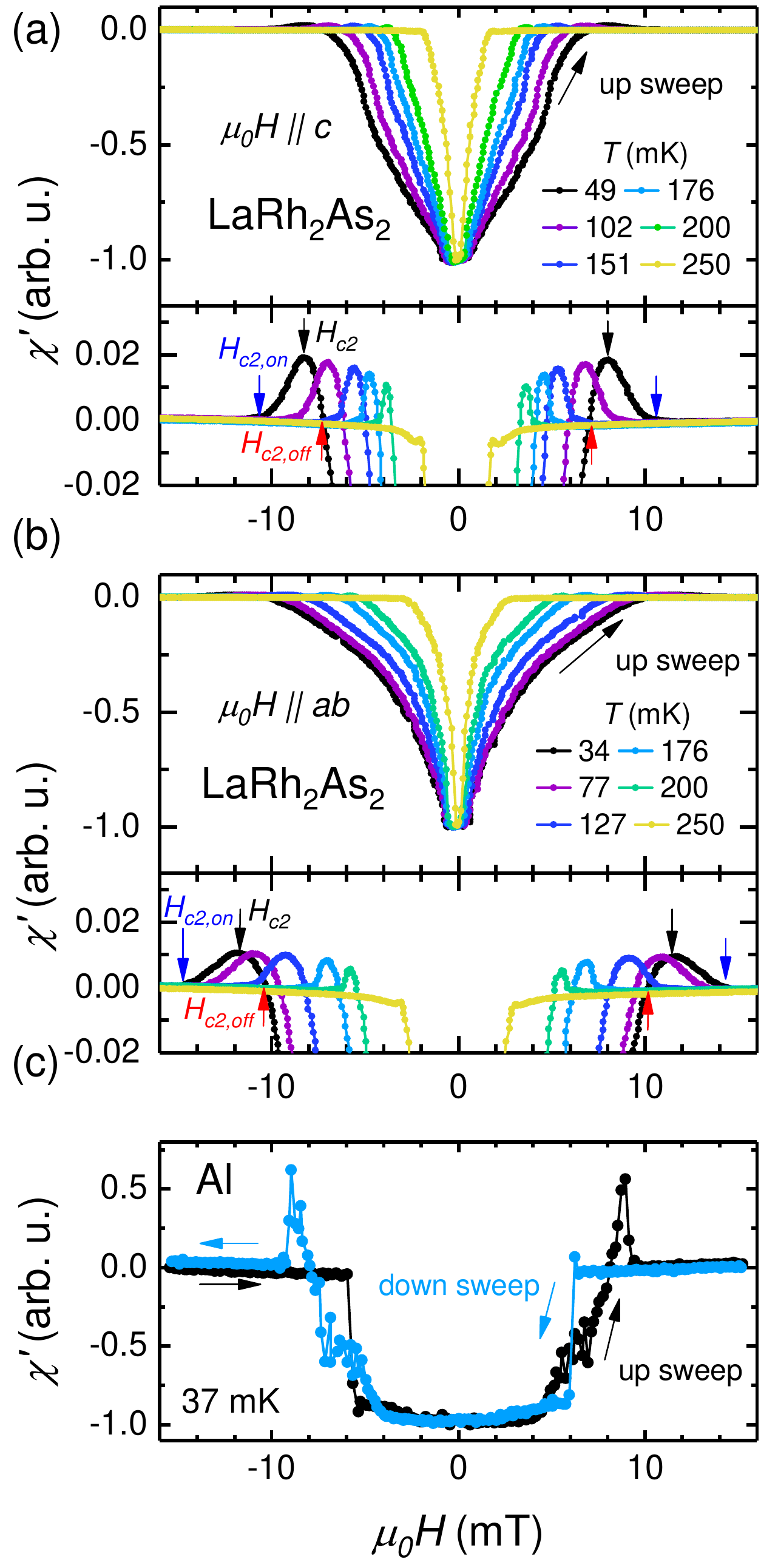}
	\caption{(a,b) Field dependence of $\chi'(H)$ for LaRh$_2$As$_2$ with $H\parallel c$ (a) and $H\parallel ab$ (b) at different temperatures. The data are obtained upon a field sweep from - 16\,mT to + 16\,mT. The lower panels emphasize the $\chi'$ peak. $H_{c2}$ is defined at the peak position. $H_{c2,on}$ ($H_{c2,off}$) denotes the onset of the peak on the high(low)-field side. (c) $\chi'(H)$ of an aluminum sample measured at 37\,mK for up and down sweeps. Aluminum exhibits the typical asymmetric differential paramagnetic effect (DPE) for a type-I superconductor.}
	\label{fig:Figure_4}
\end{figure}

\subsection{Superconducting phase diagram}

Figure \ref{fig:Figure_4}(a) and (b) show the field-dependent $\chi'$ under two different field directions.
The superconducting state is mainly characterised by a diamagnetic signal ($\chi' = dM/dH < 0$) below the critical field and below $T_c$. The onset field of the transition decreases monotonically with increasing temperatures, consistent with the nature of superconductivity. The onset field at a given temperature is higher for $H \parallel ab$ than for $H \parallel c$.

In addition, we found a small positive peak at the transition for temperatures below 200\,mK as shown in the bottom panels of Fig. \ref{fig:Figure_4}(a,b). There is a tiny peak in $\chi'$ at 250 mK as well but it does not extend to positive values. Usually such a positive peak in $\chi'$ is regarded to be the differential paramagnetic effect (DPE) that is occasionally observed in some type-I superconductors. It is attributed to the magnetization curve as a function of increasing field in which the Meissner state abruptly disappears at a critical field \cite{Hein1961,Yonezawa2005,Zhao2012}.
The typical behavior of the DPE is demonstrated in the field-dependent $\chi'$ of aluminum ($T_c$ = 1.2 K) as shown in Fig. \ref{fig:Figure_4}(c). In aluminum, the size of the DPE peak is comparable to the diamagnetic signal.
Importantly, the DPE peak appears only at the transition from the superconducting to the normal state, on the positive (negative) field upon the up (down) sweep showing a clear irreversible character. In contrast, the $\chi'$ peak in LaRh$_2$As$_2$ is much smaller, accounting for 2 \% of the diamagnetic signal and it occurs reversibly in both the up and down-sweeps. This demonstrates that the observed $\chi'$ peak in LaRh$_2$As$_2$ is unlikely associated with the nature of type-I superconductivity.

Instead, the reversible character of the $\chi'$ peak suggests the transition between an irreversible and a reversible superconducting vortex state in a type-II superconductor \cite{BANERJEE1998}. In the irreversible state, expelling external fields leads to a negative $\chi'$ signal. In the reversible state in higher fields, the magnetization decreases in magnitude with increasing fields and the positive magnetization slope ($dM/dH>0$) results in the positive $\chi'$.
The positive $\chi'$ approaches nearly zero as the normal state is realized in further fields. This is also consistent with the absence of the peak at zero field (see the inset of Fig. \ref{fig:Figure_3}(b)) and the decreasing size upon approaching $T_c$ (the 250\,mK data in the inset of Fig. \ref{fig:Figure_4}(a)) as the reversible region gradually shrinks and vanishes when approaching $T_c$ at zero field. Within the picture of type-II superconductivity, we define the upper critical field $H_{c2}$ at the peak. Additionally we define the onset $H_{c2,on}$ and offset $H_{c2,off}$ fields as given in Fig. \ref{fig:Figure_4}(a) and (b), in between which the reversible state develops.

\begin{figure}[h]
	\centering
	\includegraphics[width=0.9\linewidth]{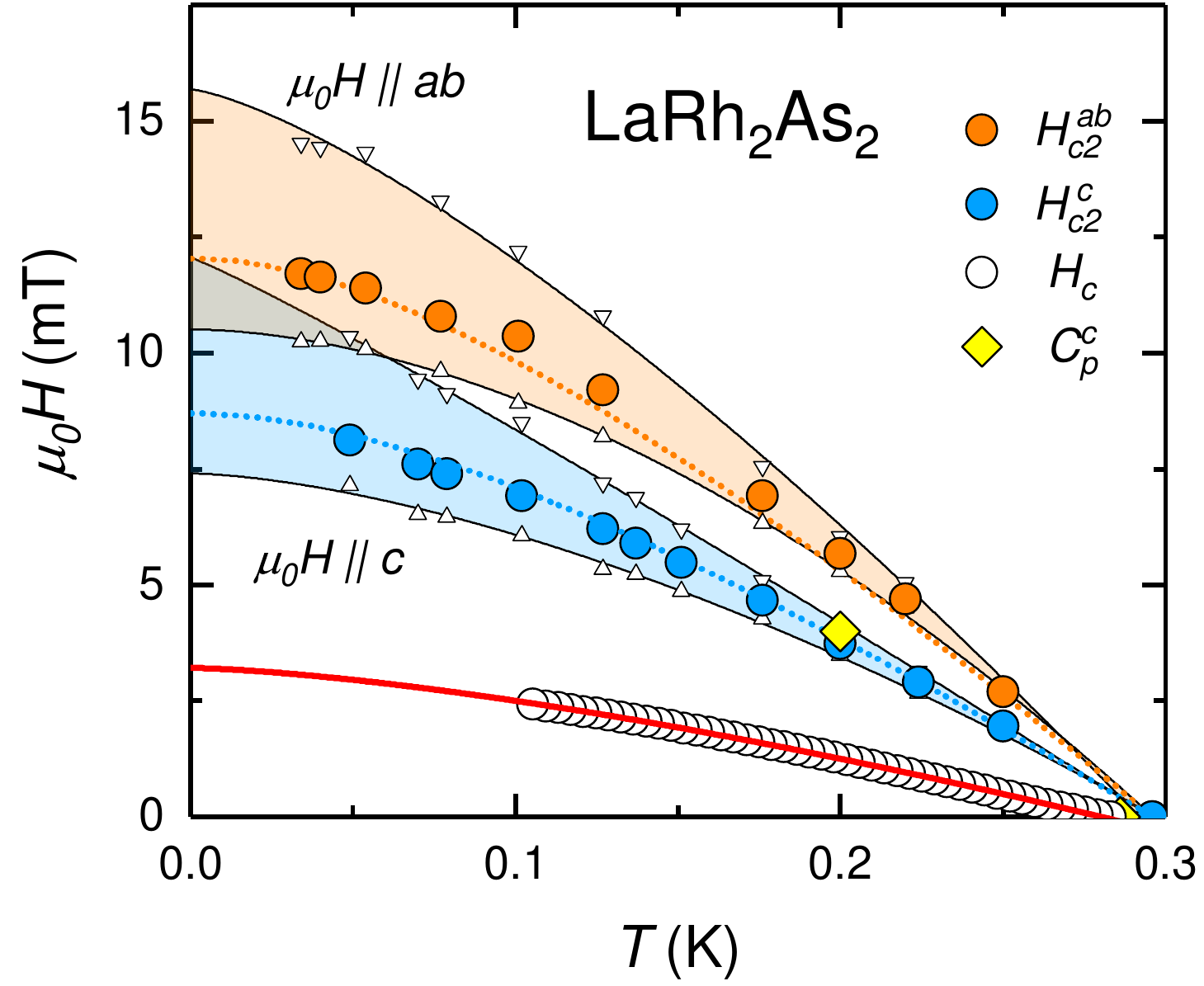}
	\caption{Superconducting phase diagram of LaRh$_2$As$_2$ for $H\parallel c$ and $H\parallel ab$. The $H_{c2}$ as well as the onset (downward triangles) and offset (upward triangles) of the superconducting transition are determined from the $\chi'$ measurements (the bottom panels of the Figure \ref{fig:Figure_4}(a,b)). The $H_{c2}$ determined from the specific-heat measurements (yellow diamond) is also shown (Fig. \ref{fig:Figure_3}(c)). The dotted lines are fits using the the dirty-limit WHH model. The black solid lines are a guides to the eye. The thermodynamic critical field $H_c$ deduced from the superconducting $C/T$ data is denoted by open circles and extrapolated to $T=0$ using $H_c=H_c(0)[1-(T/T_c)^n]$ (red line).}
\label{fig:Figure_5}
\end{figure}

The superconducting $H_{c2}$ phase diagram for $H \parallel c$ and $H \parallel ab$ is shown in Fig. \ref{fig:Figure_5}. 
The temperature dependence of $H_{c2}$ is well described by the Werthamer-Helfand-Hohenberg (WHH) model without including the Pauli limiting effect \cite{WHH1966} given by

\begin{equation}
    \ln\left(\frac{1}{t}\right)=\psi\left(\frac{1}{2}+\frac{\bar{h}}{2t}\right)-\psi\left(\frac{1}{2}\right),
\label{eq:WHH}
\end{equation}
where $\psi$ is the digamma function, $t=T/T_c$ and $\bar{h}=4H_{c2}/\pi^2(-dH/dt)_{t=1}$, with $T_c = 0.296$ K defined at the onset of the transition in $\chi'$($T$) (see Fig. \ref{fig:Figure_3}(b)). Using Eq. \ref{eq:WHH} with $dH/dT|_{T_c} = -59$ mT/K and $dH/dT|_{T_c} = -42$\,mT/K we obtain that the zero temperature limit of $H_{c2}(0))$ is 12\,mT and 8.6\,mT for $H \parallel ab$ and $H \parallel c$, respectively. The small anisotropy of $H_{c2}$ observed here indicates, that the Fermi surface of LaRh$_2$As$_2$ is rather three dimensional.

The good agreement of the $H_{c2}$ curves with the WHH model implies that superconductivity in LaRh$_2$As$_2$ is orbitally limited with a critical field that lies well below the expected BCS Pauli limit of $H_p \approx 1.84 T_c \approx 0.5$\,T. 
Therefore a possible anisotropic enhancement of the Pauli limiting field due to e.g. Rashba effect remains invisible in the superconducting phase diagram. 
Accordingly, the phase transition between even and odd parity as observed in CeRh$_2$As$_2$ \cite{khim2021} is only possible when the Pauli limit is smaller than the orbital limit which is not the case here.

\subsection{Type-II superconductivity}

We estimate the thermodynamic critical field ($H_c$) in order to extract further superconducting parameters.
$H_c$ is determined from the free-energy difference between the normal and the superconducting states, $\Delta F= F_n-F_s$, obtained from the $C/T$ data by the integration of entropy difference \cite{Bauer2004,Bauer2007,Kneidinger2013,Kawamura2018}:

\begin{equation}
  \Delta F(T)=\frac{\mu_0H^2_c(T)}{2}=\int_{T_c}^{T}\int_{T_c}^{T'}\frac{C_s-C_n}{T''}dT''dT'.
\label{eq:Hc}
\end{equation}

Here $C_n$ and $C_s$ are given by the specific heat for 10\,mT and zero field, respectively (Fig. \ref{fig:Figure_3}(c)).
The specific-heat data were integrated from $T_c$ down to 0.1\,K where the experimental data are available.
As shown in Fig.\,\ref{fig:Figure_5}, the resulting $H_c$ is smaller than the measured $H_{c2}$ with the extrapolated $H_c(0)$ of 3.2\,mT.

By using the values of $H_c$ and $H_{c2}$, we calculate the Ginzburg-Landau parameters, $\kappa_{GL}^{ab}=H_{c2\parallel c}(0)/\sqrt{2}H_c(0)=1.9$ and $\kappa_{GL}^{c}=H_{c2\parallel ab}(0)/\sqrt{2}H_c(0)=2.7$. The $\kappa_{GL}>1/\sqrt{2}$ indicates type-II superconductivity. The lower critical field, $H_{c1} =(H_c/\sqrt{2}\kappa_{GL}^{ab})\ln{(\kappa_{GL}^{ab})}$, is estimated to 0.8\,mT. This $H_{c1}$ seems to agree with the field where $\chi'$ starts to deviate from the plateau-like minimum (see the Fig. \ref{fig:Figure_4}(a)). The coherence lengths $\xi_{GL}^{ab}=200$\,nm and $\xi_{GL}^{c}=140$\,nm were estimated using $\xi_{GL}^{ab}=[\Phi_0/2\pi\mu_0H_{c2\parallel c}(0)]^{1/2}$ and $\xi_{GL}^{c}=\Phi_0/2\pi\xi_{GL}^{ab}(0)H_{c2\parallel ab}(0)$.
The penetration depths are given by the relations of $\lambda_{ab} = \kappa_{GL}^{ab}\xi_{GL}^{ab}=370$\,nm and  $\lambda_c = (\kappa^{c}_{GL})^2\xi_{GL}^{c}\xi_{GL}^{ab}/\lambda_{ab}=520$\,nm. 

In order to classify the superconductivity in terms of  dirty or clean limit, we estimated the mean free path with \cite{Orlando1979,Ramkrishnan1995}:
\begin{equation}
    l=1.27\times10^{4}\cdot[\rho_0(n^{2/3}S/S_F)]^{-1}\approx\frac{1200 \,\mu\Omega\mbox{cm}\mbox{\AA}^{-1}}{k^2_F\rho_0}.
\end{equation}
With $k_F=0.57$\AA$^{-1}$~from the DFT calculations, assuming a spherical Fermi surface ($S/S_F=1$) and using the experimental value of $\rho_0=10 \mu\Omega\mbox{cm}$, the mean free path is about $l=36$ nm. The ratio $l/\xi^{GL}_{ab}\approx0.18$ suggests that the superconductivity in LaRh$_2$As$_2$ is in the dirty limit. The fact that the $H_{c2}$ in LaRh$_2$As$_2$ is very small naturally leads to a large Ginzburg-Landau coherence length, hence, the clean limit for this material is only possible for very high purity crystals with residual resistivity at least 15 times smaller than the sample we measured. Even if we consider an uncertainty of one order magnitude in the estimation of the mean free path, the superconductivity is still within the dirty limit. Comparing with other weak-coupling non-centrosymmetric La systems, i.e., LaPt$_3$Si with comparable values of residual resistivity (2 times larger) \cite{Kneidinger2013} but with higher $H_{c2}$ (40 times larger), the superconductivity was estimated to be in the dirty limit. We can safely suggest that the superconductivity in LaRh$_2$As$_2$ is in the dirty limit.

\subsection{Estimation of electron-phonon coupling constant}

\begin{figure}[thb]
	\centering
	\includegraphics[width=0.9\linewidth]{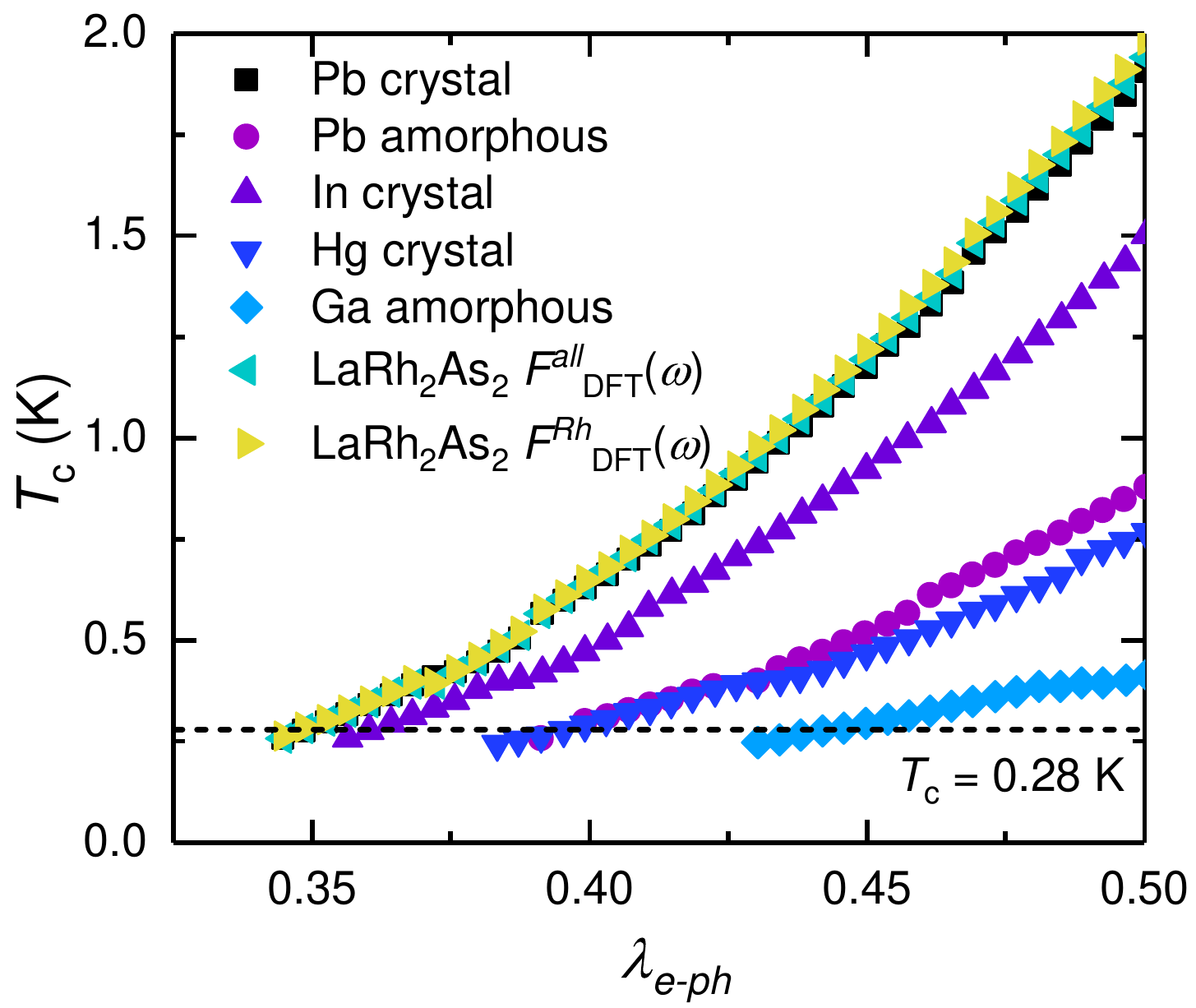}
	\caption{$T_c$ as a function of $\lambda_{e-ph}$ for various model spectra. The $\lambda_{e-ph}$ values reproducing $T_c$ = 0.28 K marked by the horizontal dotted line are listed in Table. \ref{tab:avg_freq}}
\label{fig:Figure_eliashberg}
\end{figure}

We estimate the strength of the electron-phonon ($e-ph$) interactions for $T_c$ = 0.28\,K based on the Eliashberg theory.
A measure for the strength of the $e-ph$ interaction, parameterized by $\lambda_{e-ph}$, is the mass enhancement of the quasiparticles in the normal state. According to the microscopic Eliashberg theory, the coupling constant depends on the $e-ph$ spectral function $\alpha^{2}F(\omega)$ and the weak Coulomb pseudo-potential $\mu^{*}$.
As $\alpha^{2}F(\omega)$ is unknown, we adopt several typical models $\alpha^{2}F_{model}(\omega)$ for the $e-ph$ coupling functions and estimate the range of $\lambda_{e-ph}$ required to reproduce $T_c$ for the different distributions of the spectral weights. We choose the maximal phonon frequency of $\hbar\omega_{max}$ = 36 meV in agreement with the maximal phonon frequency estimated from density functional theory (DFT) calculations. The variation with frequency of $\alpha^{2}F_{model}(\omega)$ is compared by considering the averaged frequencies

\begin{equation}
  \langle \omega \rangle_{1} = \frac{\int_{0}^{\infty}d\omega \alpha^2 F_{model}(\omega)}{\int_{0}^{\infty}d\omega \frac{\alpha^2 F_{model}(\omega)}{\omega}};   \langle \omega \rangle_{2} = \frac{\int_{0}^{\infty}d\omega \omega \alpha^2 F_{model}(\omega)}{\int_{0}^{\infty}d\omega \alpha^2 F_{model}(\omega)}.
    \label{eq:avg_freq}
\end{equation}

The model $e-ph$ coupling function is scaled by a constant prefactor so as to reproduce the observed $T_c$.
$T_c$ is calculated by solving the linearized Eliashberg equations for this scaled functions assuming a standard value for the Coulomb pseudo-potential $\mu^{*}$ $\sim$ 0.13. From the scaled $\alpha^{2}F(\omega)$, we deduce the corresponding electron-phonon mass enhancement as

\begin{equation}
\lambda_{e-ph} = 2\int_{0}^{
\infty} \frac{d\omega}{\omega}\alpha^2 F(\omega).
  \label{eq:lambda_ep}
\end{equation}

The variation of $T_c$ with $\lambda_{e-ph}$ for each model spectrum is shown in Fig. \ref{fig:Figure_eliashberg}.
$\lambda_{e-ph}$ and averaged frequencies of $\langle\omega\rangle_{1}$ and $\langle\omega\rangle_{2}$ corresponding to the experimental $T_c$ = 0.28\,K are listed in Table. \ref{tab:avg_freq}. While the estimated $\lambda_{e-ph}$ has a variation depending on the models, they are in the range of $\lambda_{e-ph}$ $\sim$ 0.34 - 0.44, consistently pointing to LaRh$_2$As$_2$ being in the weak-coupling regime.

\begin{table}[htbp]
\caption{\label{tab:avg_freq}Averaged frequencies calculated for the model spectra and the mass enhancement required to reproduce $T_c = 0.28$\,K. Although the $\lambda_{e-ph}$ have a certain spread they all point to the weak-coupling regime.}
\begin{ruledtabular}
\begin{tabular}{cccc}
& $\hbar\langle\omega\rangle_{1}$ [K] & $\hbar\langle\omega\rangle_{2}$ [K] & $\lambda_{e-ph}$\\
\hline
 Pb crystal & 181 & 206 & 0.34\\
 Pb amorphous & 130 & 194 & 0.39\\
 Hg crystal & 90 & 150 & 0.39\\
 In crystal & 158 & 201 & 0.36\\
 Ga amorphous & 82 & 168 & 0.44\\
 $F_{DFT}^{all}(\omega)$ & 191 & 223 & 0.34\\
 $F_{DFT}^{Rh}(\omega)$ & 184 & 212 & 0.34\\
\end{tabular}
\end{ruledtabular}
\end{table}

\subsection{Electronic structure calculations}

To get more insight into the electronic structure and the states relevant for the superconductivity, we carried out density functional calculations. 
Since there was no precise information about the internal coordinates of the atoms, we relaxed their position with respect to the total energy, using the experimental lattice parameters. As starting values, we used the closely related LaIr$_2$As$_2$ structure \cite{Pfannenschmidt2012}. The obtained free parameters for the different Wyckoff positions where La 2$c$ $z$=0.7544, Rh(2) 2$c$  $z$=0.1195 and As(2) 2$c$  $z$ = 0.3641. The calculated equilibrium positions vary only slightly with respect to LaIr$_2$As$_2$ (La 2$c$ $z$= 0.7550, Rh(2) 2$c$ $z$ = 0.1161 and As(2) 2$c$ $z$ = 0.3625), providing additional trust in the accuracy of the DFT calculations. 
The energy gain from the relaxation is about 1\,meV per atom, only. The corresponding differences for the valance band states are insignificant. 

\begin{figure}[tbh]
\includegraphics[width=0.48\textwidth]{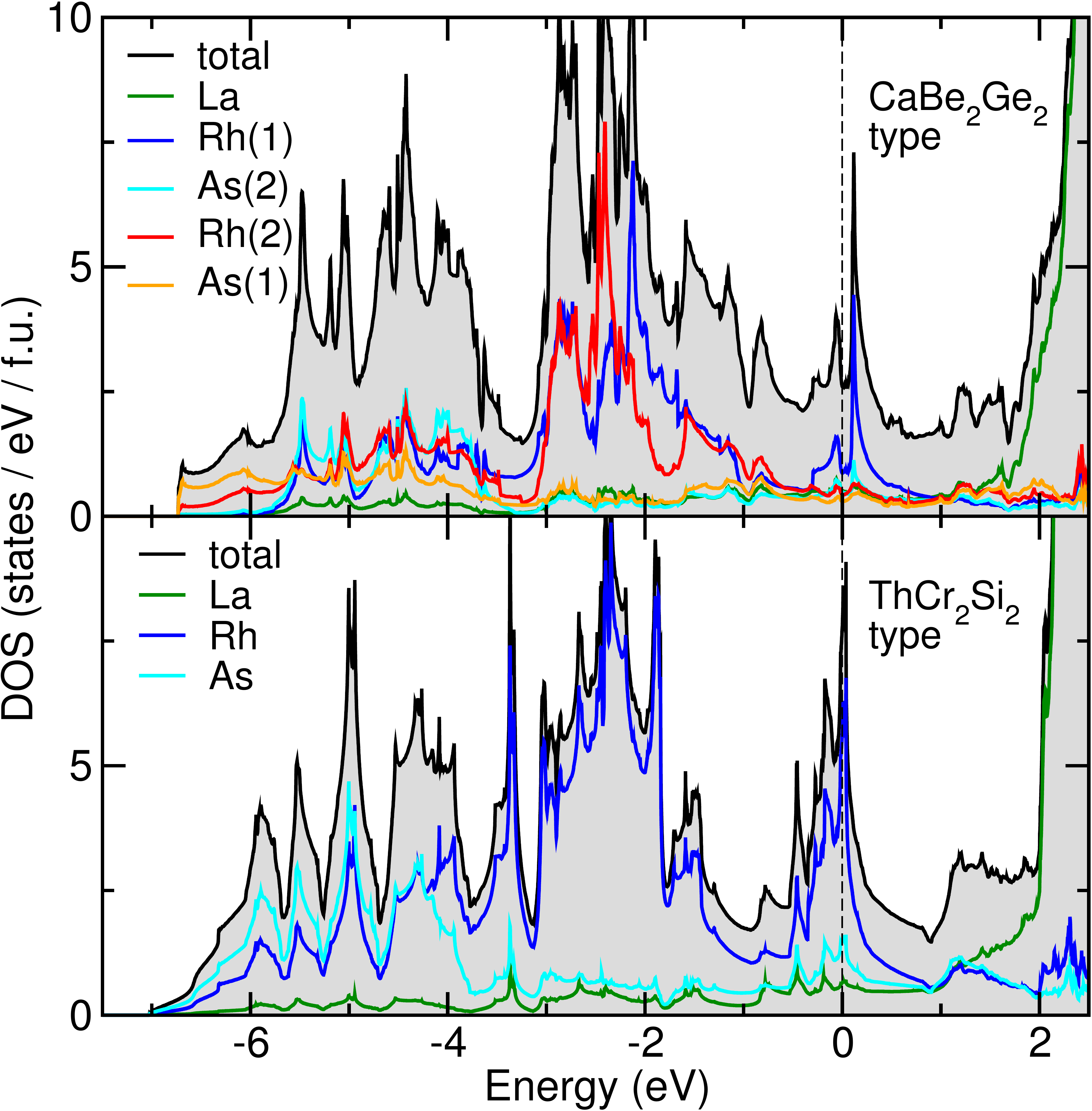}
	\caption{Total and partial electronic densities of states of LaRh$_2$As$_2$ for the experimentally observed CaBe$_2$Ge$_2$-type structure (upper panel) and the fictitious ThCr$_2$Si$_2$-type structure (lower panel). The Fermi level is at zero energy.}
\label{fig:dos}
\end{figure}

The resulting total and partial densities of states (DOS) for the CaBe$_2$Ge$_2$-type structure type (with La on the Ca site without inversion symmetry) and relaxed Wyckoff positions are shown in Fig.~\ref{fig:dos} (upper panel). The upper part of the valence band (between about -3 eV and -1 eV) is dominated by Rh states. Arsenic states contribute mostly to the bonding region of the valence band (between about -3.5 eV and -6.5 eV).

The difference between the two crystallographically different Rh-As layers is rather pronounced. The states of the Rh(1) centered layer are significantly higher in energy than those of the As(1) centered layer, in particular at the band bottom. This is a consequence of the shorter and thus stronger Rh(2)-As(1) bonds (see subsection "Crystal structure"), which in turn leads to more hybridized orbitals. 
At the Fermi level $E_F$, all atoms contribute almost equally apart from Rh(1) which exhibits a pronounced double peak feature near $E_F$. The Fermi level falls in the dip between the two peaks.

Since many of the rare-earth transition metal pnictides with a 1:2:2 stoichiometry crystallize in the ThCr$_2$Si$_2$-type structure (with the rare-earth atom on the inversion site) we calculated for comparison LaRh$_2$As$_2$ in this fictitious structure, containing only Rh centered Rh-As layers. We used the lattice parameters for the experimentally observed CaBe$_2$Ge$_2$-type structure, relaxing the 4$e$ Wyckoff position of As ($z$ = 0.3751). The resulting DOS are shown in Fig.~\ref{fig:dos} (lower panel). The DOS for the fictitious compound is similar to the real compound (upper panel), with dominating Rh states between about -3 eV and -1 eV. Since the fictitious compound contains only Rh centered Rh-As layers, the peak around the Fermi level is even more pronounced. The high value of the DOS at $E_F$ is likely the reason why the compound does not crystallize in the ThCr$_2$Si$_2$-type structure. Comparing the calculated total energies for both structures, the observed CaBe$_2$Ge$_2$-type structure is favored by 1.12 eV per formula unit. For the closely related LaNi$_2$As$_2$ compound for which both polymorphs exist \cite{Ghadraoui1988}, the calculated energy difference between them amounts to only 300 meV. Taking into account the experimental $c$/$a$ ratio reduces the difference further to 150 meV. We conclude that the large energy difference of 1.12 eV for the two polymorphs for LaRh$_2$As$_2$ makes it unlikely that the ThCr$_2$Si$_2$-type structure is stable for this compound at ambient conditions.

The calculated DOS at $E_F$ yields a bare Sommerfeld coefficient  $\gamma_0$ between 6.1 and 7.5 mJ/mol-K$^2$, depending on the choice of the exchange-correlation functional (LDA vs. GGA) or the structural input (Wyckoff positions of LaIr$_2$As$_2$ vs. relaxed positions). Taking into account the experimental $\gamma$ = 9.4 mJ/(mol-K$^2$), we can estimate a mass renormalization $\lambda = \gamma / \gamma_0 -1  = 0.4 \pm 0.14$. This compares rather well with the electron-phonon coupling constant $\lambda_{e-ph}$ $\sim$ 0.34-0.44 calculated from $\alpha^2 F$.

\begin{figure}[tbh]
\includegraphics[width=0.48\textwidth]{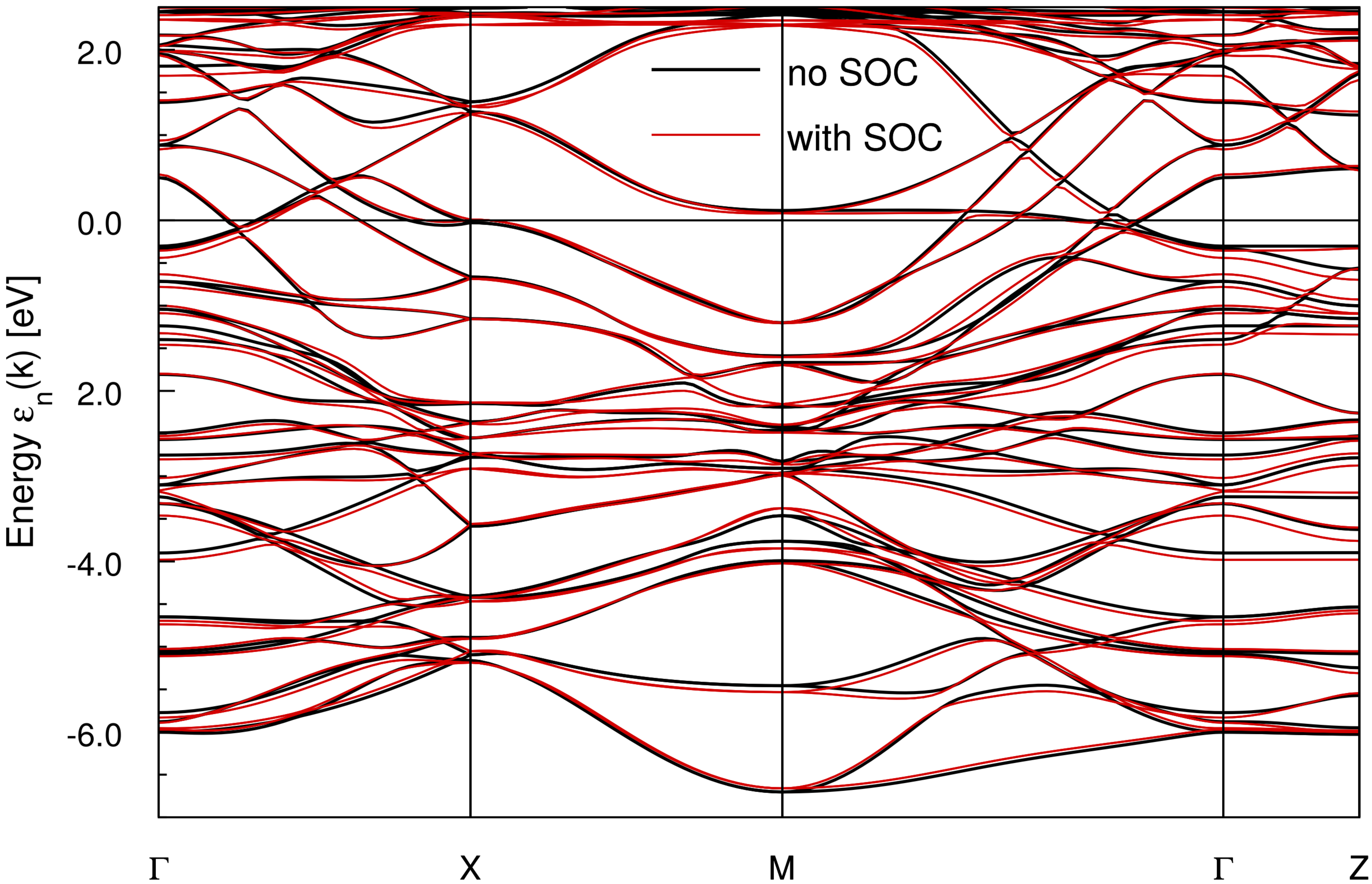}
\includegraphics[width=0.48\textwidth]{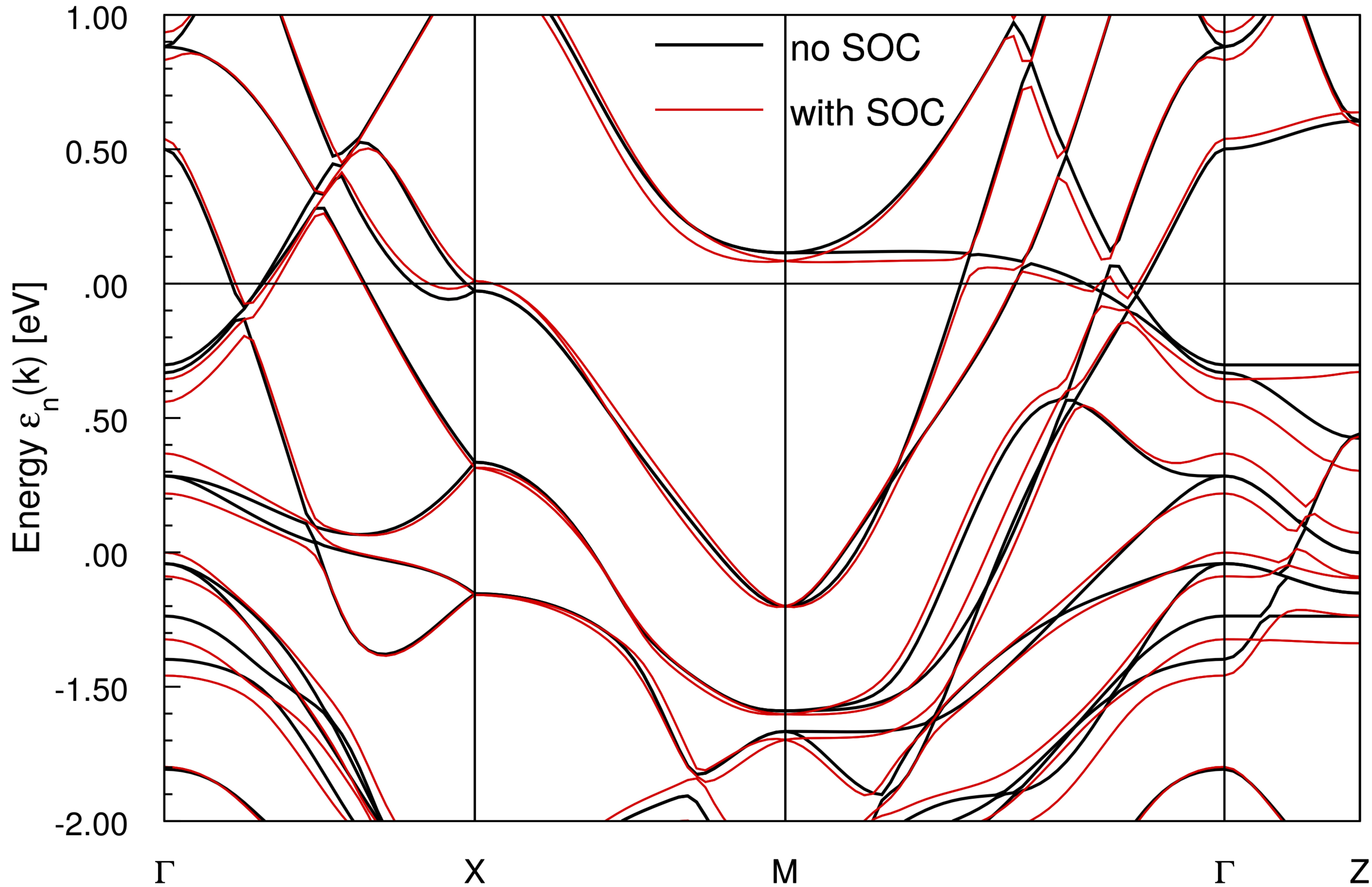}
	\caption{Band structure of LaRh$_2$As$_2$ for the experimentally observed CaBe$_2$Ge$_2$-type structure with and without spin-orbit coupling (SOC), full valence band (upper panel) and zoom-in to the Fermi level at zero energy (lower panel).}
\label{fig:bands}
\end{figure}

The band structure of LaRh$_2$As$_2$ is shown in Fig.~\ref{fig:bands}. 
Comparing the overall in-plane dispersion ($\Gamma$-X-M-$\Gamma$) with the out-of-plane-dispersion ($\Gamma$-Z), a rather pronounced two-dimensionality of the compound is observed. In addition, the influence of spin-orbit coupling (SOC) is rather small. Typical band splittings by SOC at the $\Gamma$ point are of the order of 100 to 150 meV (see Fig.~\ref{fig:bands}).

\subsection{Comparison to other (locally) non-centrosymmetric superconductors}

We also calculated the band structure of the isostructural CeRh$_2$As$_2$ compound to check whether the small differences and details in the crystal structure have a significant influence on the states near the Fermi level. 
This could be well possible because of the strongly pronounced Rh(1) peak near $E_F$. To separate the influence of the Ce-4$f$ electrons from that of the crystal structure, we treated the 4$f$ electrons as localized, non-hybridizing core states. The resulting band structure and DOS are very similar to that of  LaRh$_2$As$_2$, the Rh(1) related peak remains above E$_F$. This strongly suggests that the unconventional superconductivity in the Ce system is intimately related to the hybridization of the 4$f$ states with the valence electrons.  

For both (La and Ce) compounds, SOC originates mainly from the Rh.
However, Ce-4$f$ electrons -- as in heavy-fermion systems in general -- have very flat bands coming from the hybridisation of the $f$ electrons with the conduction electrons (here mainly the Rh $d$ electrons). Compared to the bandwidth of these flat bands, SOC is then large and has a strong effect on the electronic properties of the system. This is in line with the observation, that in non-centrosymmetric heavy-fermion superconductors, where the Ce position has the same local symmetry $C_{4v}$ as in CeRh$_2$As$_2$, unconventional superconductivity occurs, whereas the La sister compounds do not show any unconventional features of the superconducting state. For example, CeRhSi$_3$ presents a huge anisotropy of the critical field under pressure where the $T_c = 1.5$\,K is maximum, similar to CeRh$_2$As$_2$. 
In contrast, LaRhSi$_3$ has a roughly isotropic superconducting state with a $T_c$ of 0.9\,K and low critical field values \cite{Settai2008}. In penetration depth experiments, it was also found that CePt$_3$Si has an unconventional gap structure with line nodes, but LaPt$_3$Si has a conventional full gap \cite{Ribeiro2009}.\\

Let us discuss the similarity of $T_c$ in CeRh$_2$As$_2$ and LaRh$_2$As$_2$. As presented above, all the experimental evidence and calculations (Eliashberg theory) suggest that LaRh$_2$As$_2$ is an electron-phonon mediated superconductor in the weak coupling limit. No additional phase occurs that might influence the superconducting state. In contrast, CeRh$_2$As$_2$ has all the characteristics of unconventional superconductivity as in other heavy-fermion systems, where a spin (or valence)-fluctuation mechanism was proposed \cite{Mathur1998}. Firstly, the effective mass is extremely high pointing at a large density of states. This should change the $T_c$ significantly, even if all other parameters like the coupling constant and characteristic phonon energy were the same in both systems. The change of $T_c$ by a higher density of state in the Ce system is most likely compensated by changes of these other parameters so that the $T_c$ is the same in the end. Secondly, in CeRh$_2$As$_2$ an additional phase exists at 0.4\,K that was suggested to be a quadrupole-density-wave order with a complex phase diagram and unclear influence on the superconducting state \cite{Hafner2022}. Thirdly, CeRh$_2$As$_2$ shows signs of quantum criticality \cite{khim2021} and antiferromagnetic fluctuations in the normal state \cite{Kitagawa2022}. It hence seems likely that a different coupling mechanism is at play here, where the coupling constant as well as characteristic energies are different so that the matching $T_c$ is a coincidence.

LaRh$_2$As$_2$ and CeRh$_2$As$_2$ are not the only superconducting materials with the CaBe$_2$Ge$_2$-type structure. For example, LaPt$_2$Si$_2$ and SrPt$_2$As$_2$ become superconducting at 1.77\,K and 5.2\,K, respectively. Moreover, those materials have a charge density wave (CDW) transition at 122\,K and 470\,K, respectively \cite{Gupta2017,Kubo2010}. Interestingly, they show a small enhancement of $H_{c2}$ over the expectation value of the WHH theory. This behavior was attributed to a possible interplay between superconductivity and CDW. In contrast to that, other compounds, like LaPd$_2$Bi$_2$ \cite{Ganesanpotti2014}, LaPd$_2$Sb$_2$ \cite{Mu2018} and the low temperature phase of  SrPd$_2$Sb$_2$ \cite{Kase2016-2}, only exhibit superconductivity with properties within the conventional frame of electron-phonon-mediated superconductivity.
\\
Another comparable La system with crystalline structure that breaks globally inversion symmetry (CeNiC$_2$-type orthorhombic
structure with space group Amm2) is LaNiC$_2$ \cite{LEE1996}. Time-reversal symmetry was reported to be broken in the superconducting state in this compound \cite{Hillier2009}. Also, LaNiC$_2$ was suggested to possess an unconventional superconducting gap structure (nodal gap and multi gap) \cite{Bonalde_2011,Chen_2013}, and superconductivity develops in the vicinity of a magnetic quantum critical point \cite{Landaeta2017}.

The conventional superconducting properties of LaRh$_2$As$_2$ as well as other materials which lack inversion symmetry locally or globally but show no additional ordered state, suggest that breaking the inversion symmetry as the only key ingredient is not sufficient to obtain unconventional properties. Rather, it seems that the lack of inversion symmetry has to be combined with superconductivity that competes with other quantum phases such as magnetism, charge density wave, multi-polar phases in proximity to quantum critical points, in order to induce unconventional superconductivity and strong effects of antisymmetric spin-orbit coupling \cite{Landaeta2018}.
\\

\section{Conclusion and Summary}
The combination of spin-orbit coupling and absence of inversion symmetry in a compound can result in unusual properties. 
We therefore investigated the compound LaRh$_2$As$_2$ which crystallizes in the tetragonal CaBe$_2$Ge$_2$-type structure where inversion symmetry is present at the global level but absent in the La layers. 
Resistivity, specific heat and ac-susceptibility measurements prove bulk superconductivity below a bulk $T_c = 0.28$\,K. 
The $H$-$T$ superconducting phase diagrams show orbitally limited $H_{c2}$ behavior for fields both parallel and perpendicular to the tetragonal $c$ axis, with a rather weak anisotropy.
The $H_{c2}$ are small, only a factor of about 3 above the thermodynamic critical field, thus the corresponding GL coherence length  is comparably longer than an estimated mean free path, pointng to type-II superconductivity.

Both an analysis based on the Eliashberg theory and a comparison between the theoretically calculated and the experimentally determined Sommerfeld coefficient indicate the electron-phonon coupling constant to be of the order of 0.4, implying weak coupling superconductivity.
The density functional calculations further reveal that the electronic density of states at the Fermi level is predominantly due to the Rh(1) atoms from the Rh centered Rh-As layer. 
The much lower contribution of Rh(2) atoms results in the CaBe$_2$Ge$_2$-type structure being energetically favored in comparison to the ThCr$_2$Si$_2$-type structure.

Despite having the same crystal structure and similar ASOC as the Ce analogue, LaRh$_2$As$_2$ appears as a conventional superconductor. The critical fields are entirely given by the orbital limit which is well below the expected Pauli limiting field. It remains an open question, whether a higher orbital limit from large effective masses would be sufficient to give rise to similar two-phase superconductivity driven by Pauli physics, as in CeRh$_2$As$_2$, or whether additional degrees of freedom and/or additional interactions connected with the 4$f$ electrons are needed.

\begin{acknowledgments}

This work has been supported by the joint Agence National de la Recherche (ANR) and DFG program
Fermi-NESt through grants GE602/4-1 (CG) and ZW77/5-1 (GZ). JL, SZ and EH acknowledge funding from the Max Planck Research Group "Physics of Unconventional Metals and Superconductors" by the Max Planck Society.
\end{acknowledgments}

%

\end{document}